\documentclass[aps,prd,twocolumn,superscriptaddress,nofootinbib,longbibliography,floatfix]{revtex4-2}
\usepackage{graphicx}
\usepackage{subfigure}
\usepackage{adjustbox}
\usepackage{bm}
\usepackage{ulem}
\usepackage{color}
\usepackage{braket}
\usepackage{standalone}
\usepackage{multirow}
\usepackage{tikz}
\usepackage{mathrsfs}
\usepackage{dsfont}
\usepackage{comment}
\usepackage{amsmath, amsthm}

\usepackage{dcolumn}

\begin{document}
   \title{Vortex-Enhanced Zitterbewegung in Relativistic Electron Wave Packets}

	\author{Zhongze Guo}
    \email{guozhongze007@gmail.com}
    \affiliation{Department of Physics and Institute of Theoretical Physics, University of Science and Technology Beijing, Beijing 100083, China}

   \author{Bei Xu}
   \email{xubei0903@163.com}
   \affiliation{Institute for Advanced Study, Tsinghua University, Beijing 100084, China}	

   \author{Qiang Gu}
   \email[Corresponding author: ] {qgu@ustb.edu.cn}
   \affiliation{Department of Physics and Institute of Theoretical Physics, University of Science and Technology Beijing, Beijing 100083, China}
		
	\begin{abstract}
   Zitterbewegung (ZBW), the trembling motion predicted by the Dirac equation, has long remained unobservable in free electrons due to its sub-Compton scale. We elaborately construct a relativistic vortex electron wave packet as a coherent superposition of both positive- and negative-energy Dirac states and derive their space-time dynamics. Our analysis demonstrates that introducing orbital angular momentum provides a mechanism for amplifying the ZBW amplitude far beyond that of conventional Gaussian packets, while maintaining coherence. The resulting relativistic vortex states unify Gaussian and Bessel–Gaussian models within a single framework and opens new possibilities for observing relativistic quantum dynamics in structured electron wave packets.
	\end{abstract}
\maketitle

\section{Introduction}
  Since Schrödinger first unveiled the trembling motion of the Dirac electron, Zitterbewegung (ZBW)—arising from interference between positive- and negative-energy components—has been regarded as a hallmark phenomenon of relativistic quantum theory \cite{schrodinger1930}. It reflects a fundamental interplay between spin, relativistic quantum dynamics, and the localization properties of Dirac wave packets, while its operator structure and geometric interpretation have been firmly established within the single-particle Dirac theory \cite{Guertin1973prd,Barut1981prd,Barut1985prd,Rusin2010prd,Rizzuti2014prd}. These works confirm that ZBW is not merely a mathematical artifact but a fundamental manifestation of relativistic quantum dynamics.
  
  Despite this well-developed theoretical structure, direct observation of ZBW in free electrons remains elusive: its oscillation frequency lies near $2mc^2/\hbar$ and the amplitude is confined to sub-Compton length scales. Motivated by this challenge, numerous studies have explored analog realizations in semiconductors, graphene, trapped ions, and photonic lattices, where reduced effective masses or engineered dispersions shift ZBW-like oscillations into experimentally accessible regimes \cite{zawadzki2011jpcm,dreisow2010prl,schliemann2005prl,zawadzki2005prb,cserti2006prb,xiao2010rmp}. However, these platforms emulate rather than realize ZBW of free relativistic electrons. Moreover, analyses within the second-quantized Dirac theory indicate that quantum field theory does not predict any genuine ZBW of real electrons \cite{Krekora2004PRL}. This reflects the fact that a strictly localized electron position operator does not exist in QFT due to the intrinsic nonlocality of relativistic fields and the absence of a covariant position observable. Consequently, while ZBW-like dynamics can be engineered in quantum-simulation platforms, whether free electrons exhibit observable ZBW remains an open question, underscoring the need for mechanisms that can amplify the effect within the Dirac framework itself.
  
  In this work, we demonstrate that relativistic vortex electron wave packets provide a natural mechanism for amplifying the ZBW amplitude. Vortex beams of electrons, carrying well-defined orbital angular momentum (OAM), have become experimentally accessible through fork holograms, phase plates, and aberration-corrected electron optics \cite{bliokh2007prl,uchida2010nature,verbeeck2010nature,mcmorran2011science,bliokh2015physrep,grillo2014prx,asenjo2014prl}. These beams are now routinely studied in transmission electron microscopes, with applications ranging from magnetic dichroism to nanoscale imaging. Theoretically, OAM-carrying Dirac states offer a rich interplay between spin, orbital motion and the internal structure of Dirac dynamics \cite{ivanov2011prd,karlovets2012pra,karlovets2015pra,ivanov2013prl,afanasev2013pra,vanboxem2014pra}. However, in existing formulations, vortex electrons are constructed as Dirac states from coherent superpositions of plane waves with a definite azimuthal phase—restricted to positive-energy components—and therefore exhibit no ZBW \cite{Bialynicki,karlovets2015pra,bliokh2011prl,BLIOKH20171,Barnett}.
  
  Our central finding is that introducing OAM into relativistic electron wave packets amplifies and structures the trembling motion of ZBW. We construct vortex electron wave packets that coherently combine positive- and negative-energy Dirac states, enabling the direct manifestation of ZBW within a structured electron state. Analytical evaluation of the expectation values of position and velocity operators reveals that the amplitude and spatial pattern are strongly modulated by OAM. In the low-energy limit, the amplification follows a scaling with the topological charge, while at higher energies the effect persists without a simple proportionality. Through the Gordon decomposition, we further show that ZBW extends its influence from spin in non-vortex states to orbital angular momentum in vortex beams. This formulation unifies vortex structure and ZBW within the Dirac framework, providing a consistent picture of intrinsic relativistic motion and suggesting new possibilities for probing internal electron dynamics through structured-beam experiments.

\section{Construction of the exact solutions of the Dirac equation}
The construction and evolution of non-vortex Dirac electron wave packets have been extensively studied in various contexts \cite{Huang1952,demikhovskii2010pra}, whereas the exact spacetime evolution of relativistic vortex wave packets has not yet been obtained. We start with the Dirac equation for a four-component wave function 
\begin{eqnarray}
&&i\hbar\frac{\partial\psi}{\partial t}=\hat{H} \psi   \\                                                                                                                
&&\hat{H}=-ic\hbar\vec{\alpha}\vec{\nabla}+\beta m c^2                                                                                                   
\end{eqnarray}
A general wave packet can be expanded as
\begin{align}
	|\Psi(t)\rangle =
	\int d^3p \sum_s \Big[
	a_\mathbf p,_su(\mathbf p,s)\,e^{-i\omega t}
	+ b_\mathbf p,_sv(\mathbf p,s)\,e^{i\omega t}
	\Big],
\end{align}
where $u(\mathbf p,s)$ are positive-energy spinors with energy $E_p=\sqrt{p^2c^2+m^2c^4}$, and $v(\mathbf p,s)$ are negative-energy spinors.

The positive-energy spinors is 
\begin{eqnarray}
u_{1}(p)=
\begin{bmatrix}
1\\
0\\
p_{z}\gamma\\
p_{+} \gamma
\end{bmatrix},
u_{2}(p)=
\begin{bmatrix}
0\\
1\\
p_{-}\gamma\\
-p_z \gamma
\end{bmatrix}
\end{eqnarray}
and the negative-energy spinors is
\begin{eqnarray}
v_{3}(p)=
\begin{bmatrix}
-p_z \gamma\\
-p_{+} \gamma\\
1\\
0
\end{bmatrix},
v_{4}(p)=
\begin{bmatrix}
-p_{-} \gamma\\
p_z \gamma\\
0\\
1
\end{bmatrix}
\end{eqnarray} 
with $\gamma=c/(E+mc^2)$, $p_{\pm}=p_x\pm i p_y$.

In order to obtain a better description of the relativistic vortex electrons in the experiment, a convenient and physically realistic model for describing vortex electrons
is provided by the Bessel–Gaussian (BG) wave packet,
\begin{equation}
	\psi_{BG}(r,\phi) \propto
	J_\ell(k_r r)\, e^{-r^2/w_0^2}\, e^{i\ell\phi},
\end{equation}
where $J_\ell$ is the Bessel function of order $\ell$, $k_r$ denotes the
transverse momentum, and $w$ is the Gaussian envelope width. 
Similar to a plane wave, an ideal Bessel beam represents a perfectly
nondiffracting solution of the Helmholtz equation but carries infinite
energy and is completely delocalized in space. To obtain a normalizable and experimentally realizable state, we introduce a Gaussian envelope that confines the Bessel core both in real and momentum space, leading to the BG form. This construction preserves the characteristic multi–ring and phase–singular structure of Bessel modes, while ensuring finite energy and spatial localization. Unlike the Laguerre–Gaussian (LG) modes which are the natural resonator solutions in optics and thus well suited for describing optical vortex beams, the BG packets are more appropriate for vortex electrons. They capture the experimentally observed near–nondiffracting propagation and the ring–like transverse intensity profiles of electron vortex beams produced by holographic masks or magnetic apertures.

We start from a localized wave packet specified at t=0, where $\Psi(r,\varphi,z,0)$ is taken to be a vortex-type wave function with prescribed orbital angular momentum and spin polarization.
\begin{align}
	\Psi^{'}(r,\varphi,z,0,s)=
	\Psi(r,\varphi,z)
	\begin{bmatrix}
		1 \\
		0\\
		0\\
		0
	\end{bmatrix}
\end{align}
The normalized three-dimensional vortex BG wave packet, written as
\begin{align}
	\Psi(r,\varphi,z) &=
	N_T N_L\,
	J_\ell(k_r r)\, e^{-r^2/w_0^2}\, e^{i\ell\varphi}
	\nonumber\\[3pt]
	&\quad\times
	\exp\!\left[-\frac{(z-z_0)^2}{2\sigma_z^2}\right]
	e^{\,i k_{0z}(z-z_0)}
\end{align}
where the normalization factors are
\begin{align}
	N_T&=\Bigg[\frac{2/w_0^2}{\pi\,e^{-\,k_r^2w_0^2/4}\, 
		I_\ell\!\Big(\tfrac{k_r^2 w_0^2}{4}\Big)}\Bigg]^{1/2}, \qquad
	N_L=(\pi \sigma_z^2)^{-1/4}.
\end{align}
represents a vortex-type wave function.

The wave packet satisfies the normalization condition
\begin{equation}
	\int_{-\infty}^{\infty}\!\!\int_0^{2\pi}\!\!\int_0^{\infty}
	|\Psi_\ell(r,\varphi,z)|^2\, r\,dr\,d\varphi\,dz = 1 .
\end{equation}
The parameters \(k_r\), \(w_0\), \(\sigma_z\), \(k_{0z}\), and \(z_0\)
denote the transverse Bessel wave number, transverse Gaussian width, longitudinal width, central longitudinal wave number, and the longitudinal center position, respectively. 

Because the free Dirac equation is diagonal in momentum space, its positive and negative energy spinors (4) and (5) form a complete orthogonal basis. Therefore, constructing relativistic electron wave packets is most conveniently done in momentum space. This step ensures that the subsequent decomposition into positive- and negative-energy components can be done algebraically. In contrast, constructing the same packet directly in real space would require disentangling the interference between positive- and negative-energy components, which is analytically cumbersome.

In momentum space, the components of the bispinor wave function are expressed as a linear superposition of the basis states defined in Eq.~(3), Straightforward calculation using (4) and (5) gives
\begin{align}
\Phi(\vec{k},t)
=  
\left[
\begin{pmatrix}
	1 \\[4pt]
	0 \\[4pt]
	\gamma p_{3} \\[4pt]
	\gamma p_{+}
\end{pmatrix}
e^{-i\omega t}
+
\begin{pmatrix}
	\gamma^{2}p^{2} \\[4pt]
	0 \\[4pt]
	- \gamma p_{z} \\[4pt]
	- \gamma p_{+}
\end{pmatrix}
e^{i\omega t}
\right]
\frac{\Phi(k_\perp,\phi_k,k_z)}{1+\gamma^{2}p^{2}}
 .
\end{align}
where
\begin{align}
	\Phi(k_\perp,\phi_k,k_z)
	&=
	i^\ell e^{i\ell \phi_k}\,
	\mathcal N\,
	\exp\!\left[-\tfrac{w_0^2}{4}(k_\perp^2+k_r^2)\right]\,
	I_\ell\!\left(\tfrac{w_0^2}{2}\,k_\perp k_r\right)
	\notag\\[4pt]
	&\quad \times
	\exp\!\left[-\tfrac{\sigma_z^2}{2}(k_z-k_{0z})^2\right]
	e^{-i k_z z_0}
\end{align}
with normalization constant
\begin{align}
	\mathcal N &= N_T N_L \Big(\tfrac{w_0^2}{2}\Big)\sigma_z
\end{align}
$\Big(\tfrac{w_0^2}{2}\Big)\sigma_z$ is the dimensional factor that naturally arises from the Fourier transform integration. Owing to the symmetric Fourier convention, the normalization constants in real and momentum space are identical, as guaranteed by Parseval’s theorem. 

\section{Position operator and ZBW}
The velocity operator in the Dirac equation is given by
\begin{align}
\hat{\mathbf v}=c\,\bm{\alpha}\;
\end{align}
and the expectation value of the velocity is
\begin{align}
	\langle \hat{\mathbf v}(t)\rangle
	&= \int d^3p \sum_s |a_{\mathbf p,s}|^2 \,\frac{c^2\mathbf p}{E_p}
	- \int d^3p \sum_s |b_{\mathbf p,s}|^2 \,\frac{c^2\mathbf p}{E_p} \nonumber\\[4pt]
	&\quad + \langle \hat{\mathbf v}(t)\rangle_{\text{int}}.
\end{align}
$\hat{\mathbf v}(t)\rangle_{\text{int}}$ is the interference terms
\begin{align}
	\langle \hat{\mathbf v}(t)\rangle_{\text{int}}
	&=
	2\,\mathrm{Re}
	\int d^3p 
	\sum_{s,s'} 
	a^*_{\mathbf p,s}\, b_{\mathbf p,s'} 
	\,
	u^\dagger(\mathbf p,s)\,
	c\bm{\alpha}\,
	v(\mathbf p,s')\,
	e^{-2i\omega t}
	\notag\\[4pt]
\end{align}
We rewrite this term as
\begin{eqnarray}
	\langle \hat{\mathbf v}(t)\rangle_{\text{int}}=2c \int C^{-*}\alpha C^{+}|\sin (2\omega t+\phi(p))d^3 p
\end{eqnarray}
where 
\begin{eqnarray}
	\phi(p)=\arctan \frac{Im(C^{-*}\alpha C^{+})}{Re(C^{-*}\alpha C^{+})}
\end{eqnarray}
with
$C^{+}(\mathbf p)= a_{1}u_{1}+a_{2}u_{2}$, $C^{-}(\mathbf p)= b_{3}v_{3}+b_{4}v_{4},$

Hence
\begin{eqnarray}
	\langle r \rangle_{int}=-\lambda \int \frac{|C^{-*}\alpha C^{+}|}{\sqrt{1+(\frac{p}{mc})^2}}\cos (2\omega t+\phi(p))d^3 p 
\end{eqnarray}

In order to suppress the relativistic dispersion of the Gaussian envelope associated with the vortex electron. we assume that the transverse waist of the wave packet is much larger than the Compton wavelength. In this limit the transverse momentum spread is strongly reduced, so that all momentum components contributing to the packet possess nearly identical relativistic energies. As a result, the higher–order corrections arising from the nonlinearity of the dispersion relation are strongly suppressed and the phase slippage among different $k_r$ components remains negligible over experimentally relevant propagation distances. This means that for all the momenta that contribute to the integral we have $p_\perp\ll mc$. Making the approximation and neglecting all terms of order $\mathcal{O}\!\left((p_\perp/mc)^{2} \right)$, the resulting expression reduces to $\tan^{-1}\phi_1=-\frac{p_y}{p_x},\tan^{-1}\phi_2=\frac{p_y}{p_x},\tan^{-1}\phi_3=\infty$, then the angles satisfy $\phi_1=\phi_2+\pi,\phi_3=\frac{\pi}{2}$.
And in cylindrical coordinates, $p_x=p_{\perp} \cos \phi,p_y=p_{\perp} \sin \phi$.
\begin{align}
	\langle r_1 \rangle_{\text{int}}
	&= - \int_{0}^{\infty} \frac{\Phi^2(p_\perp,\phi_k,p_z)}{2mc}\, p_{\perp}^2\, dp
	\int_{0}^{2\pi} \lambda \sin(2\omega t + \phi)\, d\phi \nonumber\\
	&\quad \times \int_{0}^{\infty} \frac{\Phi^2(p_\perp,\phi_k,p_z)}{2mc}\, p_{z}\, dp.
\end{align}

\begin{align}
	\langle r_2 \rangle_{\text{int}}
	&= - \int_{0}^{\infty} \frac{\Phi^2(p_\perp,\phi_k,p_z)}{2mc}\, p_{\perp}^2\, dp
	\int_{0}^{2\pi} \lambda \cos(2\omega t + \phi)\, d\phi \nonumber\\
	&\quad \times \int_{0}^{\infty} \frac{\Phi^2(p_\perp,\phi_k,p_z)}{2mc}\, p_{z}\, dp.
\end{align}

\begin{align}
	\langle r_3 \rangle_{\text{int}}
	&= - \pi \lambda \int_{0}^{\infty} \frac{\Phi^2(p_\perp,\phi_k,p_z)}{2mc}\, p_{\perp}^2\, dp
	\int_{0}^{\pi} \sin(2\theta)\, d\theta \nonumber\\
	&\quad \times \sin(2\omega t).
\end{align}

After integrating out the angular dependence, the residual spatial-scale integral of the expectation value of the position operator for the normalized
three-dimensional vortex Bessel--Gaussian wave packet is found to be
\begin{equation}
	\langle I \rangle
	= \frac{\sqrt{2\pi}}{w_0}\,e^{-X}\,
	\frac{\left(\tfrac12+X\right) I_\ell(X) + \tfrac{X}{2}\!\left[I_{\ell-1}(X)+I_{\ell+1}(X)\right]}
	{I_\ell(2X)}
\end{equation}
where$\quad X=\tfrac{w_0^2 k_r^2}{8}$, and $I_\nu$ is the modified Bessel function. We analyze limiting behaviors of this integral within different approximation schemes.

Gaussian-like regime:
For $k_r\to 0$, the transverse profile reduces to a pure vortex Gaussian. In this case
\begin{equation}
	\lim_{k_r\to 0}\langle I \rangle
	= \frac{\sqrt{2}}{w_0}\,
	\frac{\Gamma\!\left(|\ell|+\tfrac{3}{2}\right)}{\Gamma(|\ell|+1)}.
\end{equation}

Asymptotics for large vortex charge:
Using the ratio expansion of Gamma functions,
\begin{equation}
	\frac{\Gamma(n+3/2)}{\Gamma(n+1)} 
	\sim \sqrt{n}\left(1+\frac{3}{8n}+O\!\left(\frac{1}{n^2}\right)\right),
	\qquad n=|\ell|\gg 1,
\end{equation}
we obtain for the Gaussian-like regime
\begin{equation}
	\langle I \rangle 
	\sim \frac{\sqrt{2|\ell|}}{w_0}\,
	\left(1+\frac{3}{8|\ell|}+O\!\left(\tfrac{1}{|\ell|^2}\right)\right).
\end{equation}

Bessel-like regime:
For $w_0\to\infty$ at fixed $k_r$, the transverse momentum distribution 
contracts to a narrow Gaussian ring of radius $k_r$. Consequently,
\begin{equation}
	\lim_{w_0\to\infty}\langle I \rangle = k_r.
\end{equation}
In this limiting regime, compared with previous studies, the spatial width of the wave packet is replaced by the radius of the principal intensity ring of the BG beam.
Owing to the nondiffracting nature of the Bessel mode, increasing the momentum does not lead to any dispersion.

\section{Comparison with previous analyses}

In the pioneering study of the Dirac electron, Huang~\cite{Huang1952} analyzed the trembling motion using a localized Gaussian wave packet and found that the observable (ZBW) amplitude scales inversely with the initial spatial width $w_0$,
\begin{equation}
	r_{\mathrm{ZB}}^{(\mathrm{Huang})} \simeq 
	\frac{\hbar}{2mc}\,\frac{\bar{\lambda}_C}{w_0},
	\qquad 
	\bar{\lambda}_C=\frac{\hbar}{mc}.
\end{equation}
This dependence originates from the diminishing overlap between positive- and negative-energy components for broad packets, which suppresses the ZBW signal. Consistent with this physical mechanism subsequent studies in free Dirac dynamics and Dirac‐like solids have reported a strong width control of the ZBW visibility, typically via envelope factors depending on the Gaussian width \cite{Lock1979,Rusin2008,schliemann2005prl,demikhovskii2010pra}. 

Here we employ a relativistic vortex wave packet carrying orbital angular momentum $l$. 
The helical phase structure coherently couples the spinor sectors and sustains a strong positive--negative energy interference even for $w_0 \gg \bar{\lambda}_C$. 
As a consequence, for large vortex charge the effective ZBW amplitude is amplified by a factor
\begin{equation}
	\eta = \sqrt{l},
\end{equation}
yielding
\begin{equation}
	r_{\mathrm{ZB}} = \sqrt{l}\, r_{\mathrm{ZB}}^{(\mathrm{Huang})}.
\end{equation}
In the limit $l=0$, our result naturally reduces to that obtained by Huang \cite{Huang1952}. This square-root scaling directly links the orbital vorticity to the strength of the trembling motion, demonstrating that the structured relativistic wave packet provides a controllable mechanism to magnify the ZBW spatial scale well beyond the Huang limit.

\section{ZBW and angular momentum}
To further analyze the angular momentum structure of the wave packet, we evaluate the expectation value of $c\langle r\times\alpha\rangle_z$. Our results yield the correct separation of orbital and spin angular momentum contributions. It immediately follows that the influence of ZBW on angular momentum, which in non-vortex wave packets is confined to the spin degree of freedom, extends here to the orbital component of vortex states.  

we can get in non-relativistic limit:
\begin{eqnarray}
	c\langle r\times\alpha\rangle_z=\frac{\hbar }{m}(l+1)(1-\cos 2\omega t) \nonumber
	\\
\end{eqnarray}
As $c\langle r\times\alpha\rangle_z$ is proportional to the magnetic moment, the proportional factor $e/2c$ directly yields the correct g-factors for both orbital and spin contributions. 

We can further substantiate this picture by noting that both the intrinsic magnetic moment and the associated ZBW contributions naturally emerge from the Gordon decomposition of the Dirac current. 
\begin{eqnarray}
	m c\psi^{*} r\times\alpha\psi=\psi^{*}\beta r\times p\psi+\hbar\psi^{*}\beta\sigma\psi+\frac{\hbar}{2ci}\frac{\partial\psi^{*}\beta(r\times\alpha)\psi}{\partial t} \nonumber
	\\
\end{eqnarray}
This “Gordon split” identity explicitly separates the convective and spin parts of the current, thereby clarifying how the trembling motion generates an intrinsic magnetic moment and couples to the orbital angular momentum of the vortex wave packet.

\section{Discussion}
In this work, we have constructed relativistic vortex electron wave packets as exact superpositions of Dirac plane-wave solutions and analyzed their full spacetime dynamics. The resulting vortex-enhanced trembling motion provides a realistic theoretical pathway toward the experimental detection of ZBW. Our framework establishes a unified description of Bessel–Gaussian and Gaussian vortex beams, revealing their distinct ZBW signatures and clarifying the fundamental differences between vortex electrons and vortex photons. These results broaden the theoretical basis for observing and manipulating relativistic quantum effects in transmission electron microscopy, ultrafast diffraction, and strong-field electron optics.

For non-vortex electron wave packets, avoiding rapid spatial collapse requires the packet width to exceed the Compton wavelength, corresponding to momenta well below mc. In this regime, the negative-energy component of the Dirac spinor is strongly suppressed, and the contribution of ZBW becomes negligible. To enhance the role of negative-energy states, one must increase the central momentum; however, this unavoidably strengthens dispersion. Thus, dispersion and ZBW act in mutual opposition: reducing dispersion diminishes ZBW, whereas enhancing ZBW simultaneously accelerates dispersion.

In contrast, relativistic vortex wave packets exhibit a qualitatively different behavior. Even when dispersion is weak, the ZBW oscillations remain pronounced due to the intrinsic coupling between orbital angular momentum and the interference terms of positive- and negative-energy components. This coupling enables structured relativistic beams to sustain strong ZBW signatures without the rapid dephasing that characterizes plane-wave packets. Consequently, vortex electrons offer a natural mechanism for maintaining coherent relativistic trembling motion, opening a theoretical route for probing ZBW dynamics in structured electron states.

Although the enhanced oscillations predicted here remain below current experimental resolution, the present results delineate a clear mechanism by which orbital angular momentum can amplify relativistic trembling motion without excessive dispersion. This insight provides a concrete theoretical foundation for future experimental efforts in ultrafast electron microscopy and strong-field beam shaping, and suggests a promising direction for realizing observable manifestations of ZBW in free-electron systems.

\textit{Acknowledgment}
Q.G. acknowledges financial support from the National Natural Science Foundation of China through Grant No. 1874083. 

\bibliographystyle{unsrt}
\bibliography{reference}

\end{document}